\documentstyle[preprint,cite,epsf,aps]{revtex}
\tightenlines
\begin{document}
\draft
\title{\Large On spin-1 massive particles coupled to a Chern-Simons field}
\author{M. Gomes$^{\,a}$, L. C. Malacarne$^{\,b}$ and A. J. da Silva$^{\,a}$}
  \address{$^{a\,}$Instituto de F\'\i sica, USP\\
 C.P. 66318 - 05315-970, S\~ao Paulo - SP, Brazil.}
\address{$^{b\,}$Departamento de F\'\i sica, Universidade Estadual de 
Maring\'a,\\
Av. Colombo, 5790 - 87020-900, Maring\'a - PR, Brazil.}

\maketitle

\begin{abstract}
We study spin one particles interacting through a Chern-Simons field. In the
Born approximation, we calculate the two body scattering amplitude considering
three possible ways to introduce the interaction: (a) a Proca like model
minimally coupled to a Chern-Simons field, (b) the model obtained from (a)
by replacing the Proca's mass by a Chern-Simons term and (c) a complex 
Maxwell-Chern-Simons model minimally coupled to a Chern-Simons field. In
the low energy regime the results show similarities with the Aharonov-Bohm
scattering for spin 1/2 particles. We discuss the one loop renormalization
program for the Proca's model. In spite of the bad ultraviolet behavior
of the matter field propagator, we show that, up to one loop the model
is power counting renormalizable thanks to the Ward identities satisfied
by the interaction vertices.

\end{abstract}
\section{INTRODUCTION}
In the recent years much work has been devoted to the study of the
properties of the Chern-Simons (CS) field \cite{Jackiw}. This was
motivated not only by its potential applications to condensed matter
physics but also because these studies have unveiled some new and
interesting aspects of the dynamics of relativistic quantum
physics. In particular, it has been noticed that in some circumstances
the CS field plays a stabilizing role providing theories with improved
ultraviolet behavior \cite{Chen}.  However, most of these
investigations have been restricted to the cases of spinless and spin
1/2 particles.  The reasons behind this fact are the notorious
difficulties found in the conventional treatment for higher spin
fields in four dimensions.  The troublesome aspects include noncausal
propagation and lack of renormalizability.  It is certainly worthwhile
to study the interaction of a CS and spin one matter fields so that the
origin of the difficulties could be better understood and perhaps new
and safer routes could be found.  With this in mind, we would like to
present here the results of some investigations concerning the
dynamics of spin one fields interacting through a CS term.

As a first observation, we note that a free, spin one particle  of mass
$M$ can be described alternatively by the Proca Lagrangian,

\begin{equation}\label{1}
{\cal L}_P=- \frac{1}{2} F_{\mu\nu}^{\dagger} F^{\mu\nu} - M^2 
\phi^{\dagger}_{\mu} \phi_{\mu},
\end{equation}

\noindent
where $F_{\mu\nu}= \partial_\mu\phi_\nu-\partial_\nu\phi_\mu$, or by the
Maxwell-Chern-Simons (MCS) Lagrangian,

\begin{equation}\label{2}
{\cal L}_{MCS}=- \frac{1}{2} F_{\mu\nu}^{\dagger} F^{\mu\nu} + \frac{M}{2}
\epsilon^{\mu\nu\rho}[\phi^{\dagger}_{\mu}\partial_\nu\phi_\rho+ \phi_\mu
(\partial_\nu\phi_\rho)^\dagger].
\end{equation}

Whereas the first formulation encompasses both modes of spin $\pm 1$,
the latter, Eq. (\ref{2}), represents only a single mode of spin $M/|M|$.
The two formulations are not entirely inequivalent, however. The model
(\ref{2}) is equivalent to the self-dual model,
\cite{Townsend,Outros}, and this last model is like a square root of
the Proca's, \cite{Khare}. Nevertheless, such equivalence does not in
general persists whenever the models are coupled to other dynamical
fields \cite{Gomes1}. In this investigation we will study the two body
scattering amplitude for the cases of minimal coupling of (\ref{1}) to
a CS field and when in (\ref{1}) the Proca mass $M^2 \phi^{\dagger}_{\mu}
\phi^{\mu}$ is replaced by a complex  CS term. We will also consider the case
of minimal coupling of (\ref{2}) to a CS field.
Analogously to the scattering of lower spin particles we will expect
to find similarities with the Aharonov-Bohm (AB)
scattering\cite{Aharonov}. We may recall that for spinless particles
the Born approximation found in the perturbative method only agrees
with the expansion of the exact result if a contact interaction,
simulated in the field theory approach by a quartic $(\phi^{\ast} \phi)^2$
interaction, is included from the beginning\cite{Lozano,Gomes}. It is also
known that in the spin $1/2$ case no new interaction is needed, the
role of the quartic interaction being played by the magnetic Pauli
term \cite{Gomes3}.

We will pursue the investigation of the spin effects on the
perturbative AB scattering by considering spin one particles. Previous
work in this direction started either with a complex Proca field
minimally coupled to the electromagnetic field \cite{Hagen} or with a
linearized Yang Mills equation\cite{Horner}. In both approaches the AB
scattering was discussed from a first quantized viewpoint. Here
 we consider the problem from the perspective of the theory of
quantum fields, i.  e., as the low energy limit of a fully quantized
relativistic theory of spin one particles interacting through a CS
field. One advantage of such procedure is that it incorporates some
purely quantum field effects, as vacuum polarization and anomalous magnetic
momentum.

For the Proca model minimally coupled to a CS field we will discuss
the one loop renormalization program and calculate the anomalous
magnetic moment of the matter field. As we will show, thanks to the Ward
identities satisfied by the basic interaction vertices, up to one loop the
model is renormalizable, in spite of the bad behavior of the
propagator of the matter field. The MCS model, on the other hand, turns out
to be power counting renormalizable to all orders of perturbation.

Our work is organized as follows. In section II we present the polarization
vectors and Feynman rules for the models mentioned above. There, we also
study the Born approximation for the two body scattering amplitudes. In section
III we discuss in detail the one loop renormalization parts for the Proca
model and also determine the anomalous magnetic moment. A discussion of
our results is presented in section IV. The paper contains also an appendix
with details of the calculations. 

\section{Polarization vectors and Feynman rules}

As a preliminary step toward our study of the AB scattering of two spin
one particles, let us examine some of the kinematic aspects of the asymptotic
theories. First of all, being a transversal field, $\partial_\mu \phi^\mu=0$,
the Proca field described by (\ref{1}) can be expanded in plane waves as
\begin{equation}\label{3}
\phi^\mu= \frac {1}{2 \pi} \int \frac{d^2p}{2 w_p} \sum_{\lambda=1}^{2}
\epsilon^{\mu}_{\lambda}[ a_\lambda {\rm e}^{-ipx}+ b^{\dag}_{\lambda}
{\rm e}^{ipx}],
\end{equation}

\noindent
where $w_p= \sqrt{{\vec p}^2+M^2}$ and the polarization vectors
satisfy the transversality condition $p_\mu
\epsilon^{\mu}_{\lambda}=0$. A convenient choice is

\begin{equation}\label{4}
\epsilon^{\mu}_{1}= (0,\, \epsilon^{ij} \frac {{p}_j}{|\vec p|})
\qquad \qquad \epsilon^{\mu}_{2}= (\frac{|\vec p|}{M},\,\frac {w_p}{M}
\frac {{ p}^i}{|\vec p|}).
\end{equation}

\noindent
The creation operators $a^{\dag}$ and $b^{\dag}$ allow us to construct the
Fock space of the asymptotic states. In this space, we found that the spin 
part of the
angular momentum operator

\begin{equation}\label{5}
J = \int d^2 x \; \varepsilon _{ij} x^i : T^{0j}: \;, 
\end{equation}

\noindent
where $T^{\mu\nu}=F^{\dagger\mu}_{\rho} F^{\rho\nu}+F^{\mu}_{\rho} F^{\dagger
\rho\nu}- g^{\mu\nu} {\cal L}_P$, is  given by

\begin{equation}
J_S= -i \int \frac{d^2p}{2w_p}\sum_{\lambda,\lambda^\prime= 1}^{2}
\epsilon_{ij}\,\epsilon^{i}_{\lambda^\prime} (p) \epsilon^{j}_{\lambda} (p)\,
\bigl[ a^{\dagger}_{\lambda^\prime}(p)a_\lambda(p)+
b^{\dagger}_{\lambda^\prime}(p)b_\lambda(p)
\bigr ].\label{5a}
\end{equation}

In the particle's rest frame we can  see that

\begin{eqnarray}
\mid p=0,s=\pm 1 > = \frac{ a^\dagger_1 (0) \pm i a^\dagger_2
  (0)}{\sqrt{2}} \mid 0>\; 
\label{6}
\end{eqnarray}

\noindent
are eigenstates of (\ref{5a}).

As discussed in \cite{Gomes2}, in the case of the MCS model, Eq. (\ref{2}) 
 one has just one 
polarization which can be taken as  
\begin{eqnarray}
\varepsilon_\alpha (k) = \left(\varepsilon_0 (k),\varepsilon_i (k)\right)\; , 
\label{7}
\end{eqnarray}
where $\varepsilon_0 (k)= \frac{\vec{k}. \vec{\varepsilon} (0)}{\mid
  M \mid}$  and $\varepsilon_i (k)=\varepsilon_i (0)
 +  \frac{\vec{k}.\vec{\varepsilon}(0)}{|M| (w_p+|M|)}k_i$ with   
$\varepsilon^{\mu}(0)=
 \frac{1}{\sqrt{2}} \left( 0,1, i\frac{M}{\mid M \mid}\right) $ being the
polarization vector in the particle's rest frame.

Minimally coupling the Proca field to a CS field, $A^\mu$, leads us to 
the Lagrangian

\begin{equation}\label{8}
{\cal L}_P=-\frac12 G^{\dagger}_{\mu\nu} G^{\mu\nu} - M^2 \phi^{\dagger}_{\mu}
\phi^\mu  + \frac{\theta}{2} \varepsilon_{\mu\nu\rho} A^\mu 
\partial^\nu A^\rho +\frac{\lambda}{2}(\partial_\mu A^\mu)^2\; ,
\end{equation}

\noindent
where $G^{\mu\nu}= D^\mu \phi^\nu - D^\nu \phi^\mu$ and $D^\mu =
\partial^\mu - i e A^\mu$. The Feynman rules associated to the above
Lagrangian are depicted in Fig. \ref{fig1}. In the Landau gauge ($\lambda 
\rightarrow \infty$), the analytic
expressions accompanying these rules are:

CS field propagator:
\begin{equation}\label{10}
D_{\mu\nu}(k)= -\frac{1}{\theta} \epsilon_{\mu\nu\rho} \frac{k^\rho}
{k^2+i\epsilon},
\end{equation}

Matter field propagator:
\begin{equation}
\Delta^{\alpha\beta}(p)= \frac{-i}{p^2-M^2+i\epsilon} \left[ g^{\alpha\beta} 
- \frac{p^\alpha p^\beta}{M^2}\right]
\label{8a}
\end{equation}

and interaction vertices ($p$ and $p^\prime$ denote the matter field's momenta)
\begin{eqnarray}
\Gamma_1^{\mu\alpha\beta} (p,p^\prime)&=& -i e \left[ (p+p^\prime)^\mu 
g^{\alpha\beta} - p^\beta g^{\mu\alpha}- p^{\prime\alpha} g^{\mu\beta}
\right] \label{8b} \\
\Gamma_2^{\mu\nu\alpha\beta}&=& i e^2\left[ g^{\mu\beta}g^{\nu\alpha}
+ g^{\mu\alpha} g^{\nu\beta} - 2 g^{\mu\nu}g^{\alpha\beta}\right]\; .
\label{8c}
\end{eqnarray}

The above propagators and vertices obey the identities
\begin{equation}
e \frac{d\Delta_{\alpha \beta}(p)}{dp_\mu}=\Delta_{\alpha \rho}(p)
\Gamma_{1}^{\mu\rho\sigma}(p,p)\Delta_{\sigma \beta}(p),\label{8d} 
\end{equation}
\begin{equation}
e \frac{d\phantom a}{dp_\nu}\Gamma_{1}^{\mu\alpha\beta}(p,p-q)= 
\Gamma_{2}^{\mu\nu\alpha\beta}\label{8e}
\end{equation}

\noindent
and

\begin{equation}p^{\prime}_{\beta}
\Gamma_1^{\mu\alpha\beta} (0,p^\prime)=p_\alpha
\Gamma_1^{\mu\alpha\beta} (p,0)=0. \label{8f}
\end{equation}

The expressions (\ref{8d}) and (\ref{8e}) are typical of gauge
theories being similar to the ones found in scalar QED.  These
properties will be helpful to discuss the ultraviolet behavior of the
Green functions.

To make contact with the Aharonov-Bohm scattering, let us study 
the low energy approximation for the scattering of two vector
particles. We assume that in the center of mass frame the incoming
particles have momenta $p_1=(w_p,\vec p)$ and $p_2= (w_p,-\vec p)$ and
spins $s_1$ and $s_2$, respectively. We will then denote the momenta
and spins of the outgoing particles by $p_{3}= (w_p,\vec
p^{\phantom{x} \prime})$, $p_{4}= (w_p,-\vec p^{\phantom{x} \prime})$
and $s_3$, $s_4$. The energy of the incoming particle is $w_p=
\sqrt{m^2+ {\vec p}^2}$, the spins $s_i$ can either take the values
$\pm 1$ and $|\vec p|= |{\vec p}^{\phantom {x}\prime}|$.  The tree
approximation for this process is

\begin{eqnarray}
M_{fi} &=& \left[\varepsilon^*_{\beta} (p_3,s_3) 
\Gamma_1^{\mu\alpha\beta} (p_1,p_3)\varepsilon_{\alpha}
(p_1,s_1)\right]
 D_{\mu\nu}(q)
\left[\varepsilon^*_{\rho} (p_4,s_4) 
\Gamma_1^{\nu\sigma\rho} (p_2,p_4)\varepsilon_{\sigma}
(p_2,s_2)\right] \nonumber \\
&\phantom {a}&+ (p_3 \leftrightarrow p_4, s_3 \leftrightarrow s_4) \; ,
\label{9}
\end{eqnarray}

\noindent
where $q=p_1-p_3$ and

\begin{equation}\label{11}
\varepsilon^{\alpha}(p,s)= \frac{\epsilon_1^{\alpha} (p) 
+i s \epsilon_2^{\alpha} (p)   }{\sqrt{2}}\; , 
\end{equation}

\noindent
with $\epsilon_1^{\alpha} (p)$  and $\epsilon_2^{\alpha} (p)$ as in (\ref{4}),
are circularly polarized vectors. From the above expressions we can 
 verify that
the scattering amplitude vanishes unless spin is conserved, i.e., 
$s_1=s_3$ and $s_2=s_4$ or $s_1=s_4$ and $s_2=s_3$.

After expanding in powers of $|\vec p|/M$, we get, in leading order,

\begin{equation}\label{12}
M_{fi}(s,\vec{p}, \vec{p}^\prime)= \frac{4 ie^2 |M|}{\theta} e^{\frac{-i s 
\alpha}{2}}
\left[ s + 2 i \cot(\alpha) \right] \; ,
\end{equation}

\noindent
where $\alpha$ is the scattering angle and $s=s_1+s_2$ is the total spin of
the incoming particles. Similarly to the spin 1/2 case, the origin of the 
constant term in Eq. 
(\ref{12}) is a Pauli interaction between each vector particle and
the magnetic field produced by the other. In the anti-parallel case these
effects cancel each other.

Let us now consider a model  in which the mass of the vector
particles has a topological origin. In such situation, one should use
the polarization vector given in Eq. (\ref{7}). One can then envisage
two possibilities to introduce the coupling to the CS field. One could
use (\ref{8}) but with the Proca mass replaced by a topological one,
i.e.,

\begin{equation}\label{13}
 M^2 \phi^{\dagger}_{\mu} \phi^\mu \rightarrow 
 \frac{M}{2}\varepsilon^{\mu\nu\rho}\phi^\dagger_\mu \partial_\nu \phi_\rho 
+ \frac{M}{2} \varepsilon^{\mu\nu\rho}\phi_\mu 
(\partial_\nu \phi_\rho)^\dagger
\end{equation}

\noindent
or just consider the MCS field described by (\ref{2}) minimally coupled to
a CS field. Had we employed
a topological mass instead of the Proca's, the Feynman rules would be the same
unless for the propagator for the matter field which would become

\begin{equation}
\Delta_{MCS}^{\alpha\beta}= \frac{-i}{p^2-M^2+i\epsilon} \left[ 
g^{\alpha\beta} - \frac{p^\alpha p^\beta}{p^2}
 + i M \varepsilon^{\alpha\beta\rho} \frac{p_\rho}{p^2}\right]\; .
\label{13a}
\end{equation}

\noindent
As this propagator has a better ultraviolet behavior than (\ref{8a}),
the corresponding theory will be in principle renormalizable.

Using this propagator we found a scattering amplitude,

\begin{equation}\label{14}
\frac{4 ie^2 |M|}{\theta} 
\left[ s + 2 i \mbox{ cot}(\alpha) \right] \; ,
\end{equation}

\noindent
which differs from (\ref{12}) just by a phase factor. In the case of minimal
coupling we get a result which contains an additional numerical
factor, $1/4$ , in the front of (\ref{14}).  The two possible couplings
give different cross sections and, at the present, there is no way to
select a preferred one. Besides that, radiative correction should
produce diverse cross sections even in the cases associated to
(\ref{12}) and (\ref{14}) where the corresponding Lagrangians differ
only by the mass terms. This is apparent from an inspection of the
asymptotic behavior of the matter field vector propagators. In the
case of the Lagrangian with a Proca mass, the longitudinal term in the
propagator spoils
renormalizability. However, a more careful analysis, to be done in the
next section, shows that the
degree of superficial divergence is actually lowered. Taking into 
consideration this
fact we conclude that the effective degree of divergence for a
generic one loop graph $\gamma$ is
 
\begin{equation}
d(\gamma) = 3- N_{A} - \frac12 N_{\phi}\; ,
\label{15}
\end{equation}

\noindent
where $N_A$ and $N_\phi$ are the number of external lines belonging to
the CS and to the matter vector field. In spite of the improved
behavior, as established by (\ref{15}), the model still suffers from
renormalization problems due to the divergence of graphs with $N_\phi$
equal to four and six. If the corresponding counterterms are added to
(\ref{8}) then the relations (\ref{8d}-\ref{8f}) will not be able to
guarantee renormalizability even at one loop. Actually, higher order loops
will contain nonrenormalizable divergences. The model
is renormalizable only up to one loop.  Of course these comments do
not apply if the mass has a topological origin.

\section{One loop renormalization}

As we will show now, the one loop contributions to the amplitudes for the
theory defined by Eq. (\ref{8}) have an effective degree of divergence as
given in (\ref{15}).
By power counting, any one loop graph with $n_{CS}$ and $n_{\phi}$ internal
CS and matter field lines, and containing $V_1$ and $V_2$ trilinear and
quadrilinear vertices  has the degree of superficial
divergence given by

\begin{equation}
d(\gamma)= 3 -n_{CS} + V_1= 3-\frac{N_{\phi}}{2} +V_1, \label{15a}
\end{equation}

\noindent
where we used

\begin{eqnarray}
n_{CS}+n_{\phi}&=& V_1+V_2,\\
2 n_{\phi}+ N_{\phi}&=& 2 V_1+ 2 V_2,\\
2 n_{CS} + N_{CS}&=& V_1+ 2 V_2.
\end{eqnarray}

\noindent
Formula (\ref{15}) follows now from the following observations:

1. (\ref{8d}) and (\ref{8e}) imply that the sum of graphs with $N_{CS}$
external CS lines will contain the $N_{CS}^{th}$  power of the external momenta
and therefore in such sum the degree of divergence is effectively reduced by
$N_{CS}$.

2. Besides that, due to (\ref{8f}) the contraction of the longitudinal part
of the matter field propagator with the vertex $\Gamma_1$ produces a result
whose degree in the loop momentum is reduced by $V_1$.

After establishing (\ref{15}) we shall now
examine each case of
divergence, i. e.  with $d(\gamma)\geq 0$, as specified by (\ref{15}).
Notice that logarithmically divergent parts are odd in the
integration variables and therefore vanish under symmetric
integration. This implies that, up to one loop, the six point vertex
function of the matter field and the four point function with external
lines associated to two CS field and two matter fields do not generate
counterterms. However the radiative corrections to the two and to the
four point vertex function of the matter field, the vacuum
polarization and the trilinear vertex  all have $d(\gamma)> 0$.  We shall
examine separately each one of them. We will employ dimensional
regularization which, for the cases under consideration, already acts as
a renormalization. Let us begin by the self-energy contributions which in
one loop correspond to the graphs shown in Fig. \ref{fig2}. Symmetric
integration show immediately that the contributions of the graphs 2(b)
and 2(c) vanish. The amplitude associated to 2(a) is (details of this
calculation are presented in the Appendix)

\begin{eqnarray}
\Sigma^{\alpha\beta}(p)= \int \frac{d^3k}{(2 \pi)^3} 
\Gamma_1^{\nu\alpha\alpha^\prime} (p,p+k) \Delta_{\alpha^\prime \beta^\prime}
(p+k) \Gamma_1^{\mu\beta^\prime\beta} (p+k,p) D_{\mu\nu}(k).
\label{16}
\end{eqnarray}

Although the degree of divergence indicates a quartic divergence, because of  
Eq. (\ref{8f}), the above expression diverges only quadratically. We obtain
\begin{eqnarray}
\Sigma^{\alpha\beta}(p)=
 \frac{e^2}{8 \pi \theta} F(p) \varepsilon^{\alpha\beta\rho}
p_\rho\; , 
\label{17}
\end{eqnarray}
where (from now on we assume $M$ to be positive)
\begin{eqnarray}
F(p)= \frac{(3 M^2 + p^2)}{2 M p^2} \left[(M^2+p^2) 
- \frac{1}{2M\sqrt{p^2}} (M^2-p^2)^2
\mbox{ Log}\left( \frac{M+ \sqrt{p^2}}{M-\sqrt{p^2}}\right)
\right] . 
\label{18}
\end{eqnarray}

\noindent
Although finite the above expression does not vanish at $p^2=M^2$. To
secure that the physical mass is $M$ one still have to proceed a
finite renormalization. Thus we define a renormalized amplitude by
\begin{eqnarray}
\Sigma_R^{\alpha\beta}(p)&=& 
\Sigma^{\alpha\beta}(p) - \delta M \varepsilon^{\alpha\beta\rho}
p_\rho
\nonumber \\
&=& \frac{e^2}{8 \pi \theta}  F_R(p)  \varepsilon^{\alpha\beta\rho}
p_\rho \; ,
\label{19}
\end{eqnarray}

\noindent
where $\delta M  = \frac{e^2 M}{2 \pi \theta}$
and 

\begin{eqnarray}
F_R (p) =  
 \frac{(M^2-p^2)}{2M p^2}
\left[3M^2-p^2
- \frac{(M^2-p^2)(3M^2+p^2)}{2M \sqrt{p^2}}
\mbox{ Log}\left( \frac{M+ \sqrt{p^2}}{M-\sqrt{p^2}}\right)
\right]  . 
\label{20}
\end{eqnarray}

In this situation we
obtain the renormalized propagator,

\begin{eqnarray}
\Delta_R^{\alpha\beta}= 
 \frac{-i}{p^2-M^2-\Sigma^\prime (p^2)} \left[ g^{\alpha\beta} 
- \left(1-\frac{\Sigma^\prime (p^2)}{p^2}\right)\frac{p^\alpha p^\beta}{M^2}
- i\Theta (p) \varepsilon^{\alpha\beta\rho}
p_\rho \right]\; ,
\label{21}
\end{eqnarray}

\noindent
where

\begin{eqnarray}
\Sigma^\prime (p^2)=  \frac{e^2}{8 \pi \theta}\frac{p^2 F_R^2 (p)}{(p^2-M^2)} 
\hskip 2 cm \mbox{and} \qquad
\Theta (p)=  \frac{e^2}{8 \pi \theta} \frac{F_R (p)}{(p^2-M^2)} \; .
\label{22}
\end{eqnarray}

\noindent
One sees that the needed counterterm corresponds to a  CS term for the
matter field.

Let us now focus our attention on the vacuum polarization contributions
to the CS field two point function. The associated graphs have been
drawn in Fig. \ref{fig3}. We have
\begin{eqnarray}
\Pi^{\mu\nu}= \Pi_A^{\mu\nu}+\Pi_B^{\mu\nu}\; ,
\label{23}
\end{eqnarray}
with
\begin{eqnarray}
 \Pi_A^{\mu\nu}&=& \int \frac{d^3k}{(2 \pi)^3} \Gamma_1^{\mu\beta\alpha}
 (p+k,k) \Delta_{\alpha\alpha^\prime} (k) 
\Gamma_1^{\nu\alpha^\prime\beta^\prime}
 (k,p+k) \Delta_{\beta^\prime\beta}(p+k)
\nonumber \\
\Pi_B^{\mu\nu}&=&\int \frac{d^3k}{(2 \pi)^3}
 \Gamma_2^{\mu\nu\alpha\beta} 
\Delta_{\alpha\beta} (k)\; .
\label{24}
\end{eqnarray}

\noindent
After some  calculations whose details are relegated to the appendix,  we get

\begin{eqnarray}
\Pi^{\mu\nu}= \frac{i e^2}{8 \pi} \Pi(p^2) \left( g^{\mu\nu}-
    \frac{p^\mu p^\nu}{p^2} \right) \; ,
\label{25}
\end{eqnarray}

\noindent
which is transversal as required by current conservation. In the  last
expression,

\begin{eqnarray}
\Pi (p^2) = \frac{(p^2 -4 M^2)}{4 M^2} \left[ 4 {M} - (p^2 + 4
  M^2) I_0 \right] \; ,
\label{26}
\end{eqnarray}

\noindent
with

\begin{eqnarray}
I_0 = \int_0^1 dx \; \frac{1}{\sqrt{M^2 - p^2 x (1-x)}}  \; .
\label{27}
\end{eqnarray}

\noindent
For low momentum $\Pi^{\mu\nu}$ approaches the expression,

\begin{eqnarray}
\Pi^{\mu\nu}= \frac{i e^2}{6 \pi M}  \left( g^{\mu\nu}p^2-
    p^\mu p^\nu \right) \; ,
\label{28}
\end{eqnarray}

\noindent
implying that the effective low momentum Lagrangian contains a Maxwell
term, analogously to what happens in the spin 1/2 case.

The lowest order contributions to the trilinear vertex come
from the first three graphs shown in Fig. \ref{fig4}.  The
corresponding analytic expressions are respectively

\begin{eqnarray} 
\Gamma_a^{\mu\alpha\beta} (p,p^\prime)&=& \int \frac{d^3 k}{(2 \pi)^3}
\left[\Gamma_1^{\rho\alpha\alpha^\prime}(p,p+k)
 \Delta_{\alpha^\prime \rho^\prime} (p+k)\right. 
\nonumber \\
& &\left.\Gamma_1^{\mu\rho^\prime\sigma^\prime}(p+k,p^\prime +k)
\Delta_{\sigma^\prime \beta^\prime} (p^\prime+k)
 \Gamma_1^{\sigma\beta^\prime\beta}(p^\prime +k,p^\prime) D_{\sigma\rho}(k)
\right],
\label{29}
\end{eqnarray}

\begin{eqnarray}
\Gamma_b^{\mu\alpha\beta} (p,p^\prime)= \int \frac{d^3 k}{(2 \pi)^3}
\left[\Gamma_1^{\rho\alpha\alpha^\prime}(p,p+k)
 \Delta_{\alpha^\prime \beta^\prime} (p+k) 
\Gamma_2^{\mu\sigma\beta^\prime\beta}
 D_{\sigma\rho}(k)
\right]
\label{30}
\end{eqnarray}

\begin{eqnarray}
\Gamma_c^{\mu\alpha\beta} (p,p^\prime)= \int \frac{d^3 k}{(2 \pi)^3}
\left[\Gamma_2^{\mu\rho\alpha\alpha^\prime}
\Delta_{\alpha^\prime \beta^\prime} (p^\prime+k)
 \Gamma_1^{\sigma\beta^\prime\beta}(p^\prime +k,p^\prime) D_{\sigma\rho}(k)
\right]
\label{31}
\end{eqnarray}

These expressions are in general very complicated but
in the low momentum regime a  great simplification occurs. Indeed,
performing the calculations on the matter field's mass shell we get

\begin{eqnarray}
\Gamma_a^{\mu\alpha\beta} (p,p^\prime)&=& \frac{-e^3 M}{8 \pi\theta}
\left[\frac{5}{4 M^2} g^{\beta\mu} 
\varepsilon^{\alpha\sigma\rho} p_\sigma p^\prime_\rho+
\frac{5}{4 M^2} g^{\alpha\mu} 
\varepsilon^{\beta\sigma\rho} p_\sigma p^\prime_\rho
- \frac{14}{4 M^2} g^{\alpha\beta} 
\varepsilon^{\mu\sigma\rho} p_\sigma p^\prime_\rho
\right. \nonumber \\
&+&\left.
p^\beta \left(-\frac{17}{24 M^2}\varepsilon^{\alpha\mu\rho} p_\rho- 
\frac{15}{8 M^2}\varepsilon^{\alpha\mu\rho} p^\prime_\rho \right)
+ p^{\prime\alpha} \left(\frac{15}{8 M^2}\varepsilon^{\beta\mu\rho} p_\rho+ 
\frac{17}{24 M^2}\varepsilon^{\beta\mu\rho} p^\prime_\rho \right)
\right. \nonumber \\
&+&\left.
p^\mu \left(\frac{2}{3 M^2}\varepsilon^{\alpha\beta\rho} p_\rho+ 
\frac{1}{M^2}\varepsilon^{\alpha\beta\rho} p^\prime_\rho \right)
+ p^{\prime\mu} \left(\frac{1}{M^2}\varepsilon^{\alpha\beta\rho} p_\rho+ 
\frac{2}{3 M^2}\varepsilon^{\alpha\beta\rho} p^\prime_\rho \right)
\right],
\label{32}
\end{eqnarray}

\begin{eqnarray}
\Gamma_b^{\mu\alpha\beta} (p,p^\prime)= \frac{-e^3 M}{8 \pi\theta}
\left[-2 \varepsilon^{\alpha\beta\mu}+
 \frac{4}{3 M^2} p^\beta\varepsilon^{\alpha\mu\rho} p_\rho
- \frac{2}{3 M^2} p^\mu\varepsilon^{\alpha\beta\rho} p_\rho
\right]
\label{33}
\end{eqnarray}

\begin{eqnarray}
\Gamma_c^{\mu\alpha\beta} (p,p^\prime)= \frac{-e^3 M}{8 \pi\theta}
\left[-2 \varepsilon^{\alpha\beta\mu}-
 \frac{4}{3 M^2} p^{\prime\alpha}\varepsilon^{\alpha\mu\rho} p^{\prime}_{\rho}
- \frac{2}{3 M^2} p^{\prime\mu}\varepsilon^{\alpha\beta\rho} p^{\prime}_{\rho}
\right].
\label{34}
\end{eqnarray}

It can be easily verified that these results satisfy current conservation
as expressed in the Ward identity

\begin{eqnarray}
e\frac{d\Sigma^{\alpha\beta} (p)}{d(p)^{\mu}}\Bigr |_{p^2=M^2} =
\Gamma_a^{\mu\alpha\beta} (p,p) +\Gamma_b^{\mu\alpha\beta} (p,p)+
\Gamma_c^{\mu\alpha\beta} (p,p) \; .
\label{35}
\end{eqnarray}

We are now in a position to calculate the vector meson anomalous
magnetic moment. Usually this done by coupling the matter field to an
external electromagnetic field. Besides that, to disentangle the
various contributions, here it is also convenient to do the
calculations in a particular frame where $\vec p 
+{\vec p}^{\phantom a\prime}=0$,
the Breit frame.  It is then found that the magnetic moment is given by
\cite{Kim}

\begin{eqnarray}
\mu = \lim_{q\rightarrow 0} \; \frac{1}{2 M}\epsilon_{ij} \frac{q^j}{q^2}
 \Gamma^i
(\vec{p}, -\vec{p})\; ,
\label{36}
\end{eqnarray}
where

\begin{eqnarray}
 \Gamma^i (\vec{p}, -\vec{p})= \varepsilon^*_\beta ( -\vec{p},s)
 \Gamma^{i\alpha\beta} (\vec{p}, -\vec{p})  \varepsilon_\alpha
 (\vec{p},s) \; ,
\label{37}
\end{eqnarray}

\noindent
and the limit prescription singles out the term proportional to $\vec q$.
In the tree approximation we get

\begin{eqnarray}
\mu = \frac{e}{2 M} g s = \pm  \frac{e}{2 M} \; , 
\label{38}
\end{eqnarray}

\noindent
where $s=\pm 1$ is the spin. The above result means that the gyromagnetic
factor is $g=1$.  The gyromagnetic factor may  be changed (sometimes
$g=2$ is desirable \cite{Glashow})  if the magnetic term \cite{Schwinger}

\begin{eqnarray}
{\cal L}_{mag} = -ie \gamma (\partial_\mu A_\nu - \partial_\nu A_\mu) 
\phi^{*\mu} \phi^\nu\;  
\label{39}
\end{eqnarray}

\noindent
is added to (\ref{8}). In such case, it is easy to check that the
magnetic moment for a spin one particle becomes $ e (1+\gamma)/(2M)$.
Nevertheless, as the new vertex (\ref{39}) does not obey Eq. (\ref{8f})
the ultraviolet behavior of the Green functions is definitely wrecked.
Thus, if one insists in having $g=2$ another approach becomes
mandatory. For this reason we shall not anymore consider the
possibility of adding (\ref{39}). After these remarks let us proceed
toward the computation of the anomalous magnetic moment. From the
expressions for the three graphs given above we have

\begin{equation}
\Gamma^{i}_{a-c}(\vec p, -\vec p)= \varepsilon^{*}_{\beta}(-\vec p, s) 
\Gamma^{i\alpha\beta}_{a+b+c} \varepsilon_\alpha (\vec p, s)=\frac{3 e^3}
{8\pi \theta}\epsilon^{ij}q_j.\label{39a}
\end{equation}

We still have to add the
contributions of the graphs 4(d) and 4(e) which corresponds to the
analytic expressions

\begin{eqnarray}
\Gamma_d^{\mu\alpha\beta} (p,p^\prime)=
\left[ \Sigma_R^{\alpha\alpha^\prime} (p)
\Delta_{\alpha^\prime \beta^\prime} (p) 
\Gamma_1^{\mu\beta^\prime\beta} (p,p^\prime)
\right]
\label{40}
\end{eqnarray}

and
\begin{eqnarray}
\Gamma_e^{\mu\alpha\beta} (p,p^\prime)=
\left[\Gamma_1^{\mu\alpha\alpha^\prime} (p,p^\prime) 
\Delta_{\alpha^\prime \beta^\prime} (p^\prime)
 \Sigma_R^{\beta^\prime\beta} (p^\prime)
\right],
\label{41}
\end{eqnarray}

However, for small momenta and in the Breit frame,
\begin{eqnarray}
\Gamma_d^{i\alpha\beta} (\vec p,-{\vec p})&=& \frac{-e^3 M}{8 \pi\theta}
\left[-\frac{1}{M^2} p^\beta\varepsilon^{\alpha i \rho} p_\rho+
\frac{1}{M^2} g^{i\beta}\varepsilon^{\alpha\sigma\rho}p_\sigma p^\prime_\rho
\right ]\label{42} \\
\Gamma_e^{i\alpha\beta} (\vec p,-{\vec p})&=& \frac{-e^3 M}{8 \pi\theta}
\left[\frac{1}{M^2} p^\alpha\varepsilon^{\beta i \rho} p_\rho+
\frac{1}{M^2} g^{i\alpha}\varepsilon^{\beta\sigma\rho}p_\sigma p^\prime_\rho
\right ]\label{43},
\end{eqnarray}

\noindent
so that, saturating with external polarizations, we get

\begin{equation}
\varepsilon^*_\beta ( -\vec{p},s)\Bigl [
 \Gamma_d^{i\alpha\beta} (\vec{p}, -\vec{p}) +\Gamma_e^{i\alpha\beta} 
(\vec{p}, -\vec{p})\Bigr ] \varepsilon_\alpha(\vec{p},s) =-\frac{2 e^3}{8\pi 
\theta}
\epsilon^{ij}q_j \label{44}.
\end{equation}

The sum of Eqs (\ref{39a}) and (\ref{44}) gives

\begin{equation}
\Gamma^i(\vec{p},-\vec{p})= \frac{e^3}{8 \pi \theta}
 \;\varepsilon^{ij} q_j.\label{45}
\end{equation}
Thus, up to one loop, the magnetic moment for vector particles of spin 
$\pm 1$ is

\begin{eqnarray}
\mu=\pm \frac{e}{2 M} \left[ 1 \pm \frac{e^2}{8 \pi \theta}\right]\; .
\label{46}
\end{eqnarray}

\noindent
It should be noticed that  the graphs with 
self energy corrections contribute significantly to the final result.
This is similar to what happens for the case of spin 1/2 particles in
the Coulomb gauge and results from the fact that in both cases the
interaction with the CS field modifies the free propagators in
an essential way. In our case, as can be seen from (\ref{17}) a CS
term is produced.

It is also worthwhile to remark that, as in the spin 1/2 case, the
anomalous magnetic moment has the same sign for both spin up and spin
down situations \cite{Gomes3}.

\section{Discussion}

The calculation of the radiative corrections to the propagators and vertices
done in the previous section allow us to incorporate vacuum polarization and
anomalous magnetic moment effects in the two body scattering amplitudes
computed earlier. In fact, in the low energy regime we get

\begin{eqnarray}
{\cal M}_{a}^{s_1=1,s_2=1,s_3,s_4} &=& \frac{ie^4M}{8 \pi \theta^2} 
(\cos\alpha -i\sin\alpha) \left\{ \frac{1+s_4}{1-\cos\alpha} 
\right. \nonumber \\
& &\left.\left[-2+ (1-s_3)\cos\alpha + (1+s_3)\cos^2\alpha -i (1+s_3)
\sin\alpha\cos\alpha \right]\frac{}{} \right\}
\label{47}
\end{eqnarray}
\begin{eqnarray}
{\cal M}_{a}^{s_1=-1,s_2=-1,s_3,s_4} &=& \frac{ie^4M}{8 \pi \theta^2} 
(\cos\alpha +i\sin\alpha) \left\{ \frac{-1+s_4}{1-\cos\alpha} 
\right. \nonumber \\
& &\left.\left[2- (1+s_3)\cos\alpha -(1-s_3)\cos^2\alpha -i (1-s_3)\sin\alpha
  \cos\alpha \right]\frac{}{} \right\}
\label{48}
\end{eqnarray}
\begin{eqnarray}
{\cal M}_{a}^{s_1=1,s_2=-1,s_3,s_4} &=& \frac{ie^4M}{8 \pi \theta^2} 
 \left\{ \frac{-1+s_4}{1-\cos\alpha} \left[(1+s_3) \right.\right. \nonumber \\
& &\left.\left. -2 s_3\cos\alpha - (1-s_3)\cos^2\alpha 
+ 2 i \sin\alpha -i (1-s_3)\sin\alpha
  \cos\alpha \right]\frac{}{} \right\}
\label{49}
\end{eqnarray}
\begin{eqnarray}
{\cal M}_{a}^{s_1=-1,s_2=1,s_3,s_4} &=& \frac{ie^4M}{8 \pi \theta^2} 
 \left\{ \frac{1+s_4}{1-\cos\alpha} 
\left[(-1+s_3) \right.\right. \nonumber \\
& &\left.\left. -2 s_3\cos\alpha + (1+s_3)\cos^2\alpha 
+ 2 i \sin\alpha -i (1+s_3)\sin\alpha
  \cos\alpha \right]\frac{}{} \right\}\; ,
\label{50}
\end{eqnarray}  

\noindent
for the contributions from the vertex corrections indicated in Fig.
5a.  The result for Fig. 5b is obtained from the above expressions by
exchanging $s_3$ and $s_4$. For the vacuum polarization diagram, Fig.
5c, we found the result

\begin{eqnarray}
{\cal M}_{c}^{s_1,s_2,s_3,s_4} &=& \frac{2ie^4M}{3\pi\theta^2} 
[(1+s_1s_3)\cos \alpha - i (s_1+s_3)\sin\alpha ]\nonumber \\
&\phantom a& [(1+s_2s_4)\cos \alpha - i 
(s_2+s_4)\sin\alpha ].
\end{eqnarray}

The next diagram, Fig. 5d, does not contribute at leading order of 
$|\vec p|/M$. To complete
the one loop calculation it is still necessary to compute the contributions
of the triangle (Fig. 5e and 5f) and box (Fig. 5g and 5h) diagrams. These
graphs present a logarithmic ultraviolet divergence (by Lorentz covariance 
the would be linear divergence does not appear). Besides that they  are
so much intricate that even using dimensional regularization and employing
an algebraic computer program the integrals did  not turned out to be feasible.
 
To summarize, in this work we have studied some properties  
of spin one particles interacting through a CS field. As argued in the
introduction, this is a complex system but we found some simplifications
which allowed us to study its behavior up to the one loop level. We
have indicated other possible scheme as the use of the complex MCS model
minimally coupled to a CS term. Although this model is power counting 
renormalizable, the algebraic structure is very much cumbersome that
no practical results are possible even at the one loop level. In such
situation the scheme adopted by us seems to be the most useful although
not being easily generalizable to higher orders. 

\begin{center}
ACKNOWLEDGMENTS  
\end{center}

This work was supported in part by Conselho Nacional de
Desenvolvimento Cient\'\i fico e Tecnol\'ogico (CNPq), Coordena\c
c\~ao de Aperfei\c coamento de Pessoal de N\'{\i}vel Superior (CAPES) and
Funda\c c\~ao de Amparo \`a Pesquisa do Estado de S\~ao Paulo (FAPESP).

\appendix
\section{} 
\label{appendix}

In this appendix we shall present some details of the calculations of the
vector meson self-energy and of the CS polarization tensor. Let us
begin considering Eq. (\ref{16}). There are two denominators so that
employing Feynman's trick
\begin{equation}
\frac{1}{AB} =\int_0^1 dx \frac{1}{[(A-B) x + B]^2}
\label{s2}
\end{equation}
and changing the integration variable, $k\rightarrow k+p x$ one finds
\begin{equation}
\Sigma^{\alpha\beta}(p)= \frac{-ie^2}{\theta}
\int_0^1 dx \int \frac{d^3k}{(2 \pi)^3} 
\frac{L^{\alpha\beta}}{[k^2 - C^2]^2}\; ,
\label{s3}
\end{equation}
where $C^2= M^2(1-x) -p^2 x(1-x)$ and
\begin{eqnarray}
L^{\alpha\beta} &=& \left\{ [3 M^2 + x p^2] (\varepsilon^{\sigma\beta\rho} k_\sigma
  k^\alpha + \varepsilon^{\alpha\sigma\rho} k_\sigma k^\beta ) p_\rho
  + [(1-x)M^2 p^2 
 \right. \nonumber \\
&-&\left. (1-x) (p_\mu k^\mu)^2  (x^2-x^3) p^4] 
\varepsilon^{\alpha\beta\sigma}p_\sigma +
2 x p^2\varepsilon^{\alpha\beta\rho}k_\rho k_\sigma p^\sigma
  \right\}/M^2\; .
\label{s4}
\end{eqnarray}
Using dimensional regularization one can perform the momentum integration
to get
\begin{eqnarray}
\Sigma^{\alpha\beta}(p)=\frac{e^2}{8\pi \theta M^2}
\int_0^1 dx \frac{6 M^4 + [p^4 -M^2 p^2 -6 M^4] x + [M^2 p^2-7 p^4]
  x^2 + 6 p^4 x^3}{\sqrt{M^2(1-x) -p^2 x (1-x)}} 
\varepsilon^{\alpha\beta\sigma}p_\sigma.
\nonumber \\
\label{s5}
\end{eqnarray}
The computation of the remaining parametric integration is straightforward 
and produces the result (\ref{18}).

Let us now turn to the polarization tensor. From (\ref{23}) and (\ref{24})
and proceeding analogously to what we have done before we arrive at
\begin{eqnarray}
\Pi^{\mu\nu}= 
\frac{e^2}{M^2} \int \frac{d^3k}{(2 \pi)^3}
\frac{N_{1}^{\mu\nu}}{[k^2-M^2] [(k+p)^2-M^2]}\; ,
\label{p3}
\end{eqnarray}
where
\begin{eqnarray} 
N_{1}^{\mu\nu}&=& [k^2 (p^\mu k^\nu + k^\mu p^\nu) - 2 k^\mu k^\nu k^\alpha
p_\alpha ] + [ 8 M^2 k^\mu k^\nu - (4M^2 g^{\mu\nu} 
+ p^\mu p^\nu)k^2
\nonumber \\
&+& k^\alpha p_\alpha ( 2 p^\mu k^\nu + 2 k^\mu p^\nu - 2 g^{\mu\nu}
k^\beta p_\beta) + p^2 (-3 k^\mu k^\nu +  2 g^{\mu\nu} k^2)]
\nonumber \\
&+& [ 3 M^2 (p^\mu k^\nu+ k^\mu p^\nu - 6  g^{\mu\nu} k^\alpha
p_\alpha)]
+[ M^2 ( 4 M^2  g^{\mu\nu} + p^\mu p^\nu - 3 p^2  g^{\mu\nu})] \; .
\label{p4}
\end{eqnarray}
Employing again Feynman's formula (\ref{s2}), translating the
integration variable,$ k\rightarrow k-p x$ and deleting the terms
odd in $k$, results
\begin{eqnarray}
\Pi^{\mu\nu}= 
\frac{e^2}{M^2} \int_0^1 dx \int \frac{d^3k}{(2 \pi)^3}
\frac{N_{2}^{\mu\nu}}{[k^2-a^2]^2}\; ,
\label{p5}
\end{eqnarray} 
where, as before, $a^2=M^2 - p^2 x (1-x)$ and
\begin{eqnarray}
N_{2}^{\mu\nu}&=&
M^2 [4 M^2 g^{\mu\nu} + p^\mu p^\nu (1-6 x +8x^2) + p^2  g^{\mu\nu}
(-3 +6 x - 4 x^2)] 
\nonumber \\
&+&
k^\mu k^\nu [ 8 M^2 - 3 p^2 + 2 p^2 x ] + k^2 [ -4 M^2  g^{\mu\nu} -
p^\mu p^\nu (1+2 x) + 2 p^2  g^{\mu\nu}] 
\nonumber \\
&+& [2 k^\mu p^\nu k^\alpha p_\alpha + 2 p^\mu k^\nu  k^\alpha p_\alpha
- 2  g^{\mu\nu}( k^\alpha p_\alpha)^2] \; .
\label{p6}
\end{eqnarray}
With the help of dimensional regularization  this becomes
\begin{equation}
\Pi^{\mu\nu}= 
\frac{ie^2}{8 \pi M^2} \int_0^1 dx
 \frac{N_{3}^{\mu\nu}}{[M^2 -p^2 x (1-x)]^{1/2}}\; ,
\label{p7}
\end{equation}
with
\begin{eqnarray}
N_{3}^{\mu\nu}&=&M^2 (2-12 x + 8x^2) [p^\mu p^\nu - p^2 g^{\mu\nu}] 
\nonumber \\
&+& p^2 [p^\mu p^\nu ( -x + 7 x^2 - 6 x^3) + p^2 g^{\mu\nu} (-x -x^2 +
2 x^3)]\; . 
\label{p8}
\end{eqnarray}
The parametric integration can be done exactly producing the result
(\ref{25}).

\begin{figure}
\caption{Feynman rules for the Proca model minimally coupled to a CS field.} 
\label{fig1}
\end{figure}

\begin{figure}
\caption{Graphs contributing to the one-loop correction to the vector meson
propagator.} 
\label{fig2}
\end{figure}

\begin{figure}
\caption{One loop vacuum polarization graphs associated to the CS field.} 
\label{fig3}
\end{figure}

\begin{figure}
\caption{Graphs contributing to the vector meson anomalous magnetic moment.} 
\label{fig4}
\end{figure}

\begin{figure}
\caption{One loop contributions to the vector meson scattering.} 
\label{fig5}
\end{figure}

\end{document}